# Is there a linewidth theory for semiconductor lasers?

**Boris Spivak**
*Dept. of Physics, University of Washington, Seattle, WA, 98195*

**Serge Luryi**
*Dept. of Electrical and Computer Engineering, SUNY at Stony Brook, Stony Brook, NY, 11794-2350*

## 1. Introduction

Laser linewidth theory was pioneered by Schawlow and Townes [1] and further developed in [2] and [3]. We discuss the status of the Schawlow-Townes-Lax-Henry (STLH) theory of laser linewidth in the instance of semiconductor injection lasers. At injection levels $I$ below threshold ($I < I_C$) one can introduce two spectra $g(\omega, I)$ and $\sigma(\omega)$, see Fig. 1a, describing, respectively, the material gain and the loss at cavity mirrors of the electromagnetic field intensity. The gain $g(\omega, I)$ is generally an increasing function of $I$. At $I = I_C$ the two spectra touch each other, $g(\omega_0, I_C) = \sigma(\omega_0)$, and the generation begins. The STLH theory of laser linewidth is based on the assumption that in the mean-field approximation (i.e., without fluctuations) the singular in frequency nature of generation persists above the threshold (i.e., for $I > I_C$). In the framework of this approach, the laser line acquires a finite width $\Gamma$ entirely due to fluctuations. In an ideal laser these fluctuations are due to the random discrete nature of spontaneous emission.

We shall refer to the property of the two spectral curves $g(\omega, I)$ and $\sigma(\omega)$ to touch each other at a singular frequency for $I > I_C$ as *rigidity*, see Fig. 1b. In principle, however, scenarios other than rigidity are also possible. For example, the curves may touch each other for $I > I_C$ in a finite interval of frequencies, see Fig. 1c, so that there is a finite linewidth even in the mean-field approximation. In this case, the account of fluctuations would provide only a correction. This is not an unusual situation. For example, the conventional mean-field scenario for multimode laser generation (Fig. 1d) involves oscillations at several discrete frequencies.

In this paper we examine the validity of the assumption of rigidity. In Sect. 2 we briefly review the standard STLH linewidth theory. In Sect. 3 we derive a mean-field expression for the linewidth using Boltzmann's kinetic equation for electrons and photons. In this approach the linewidth turns out to an increasing function of injection, which violates the assumption of rigidity and is in contradiction with the STLH scenario. Curiously, however, it is not necessarily in contradiction with experiment, see the discussion in Sect. 4.

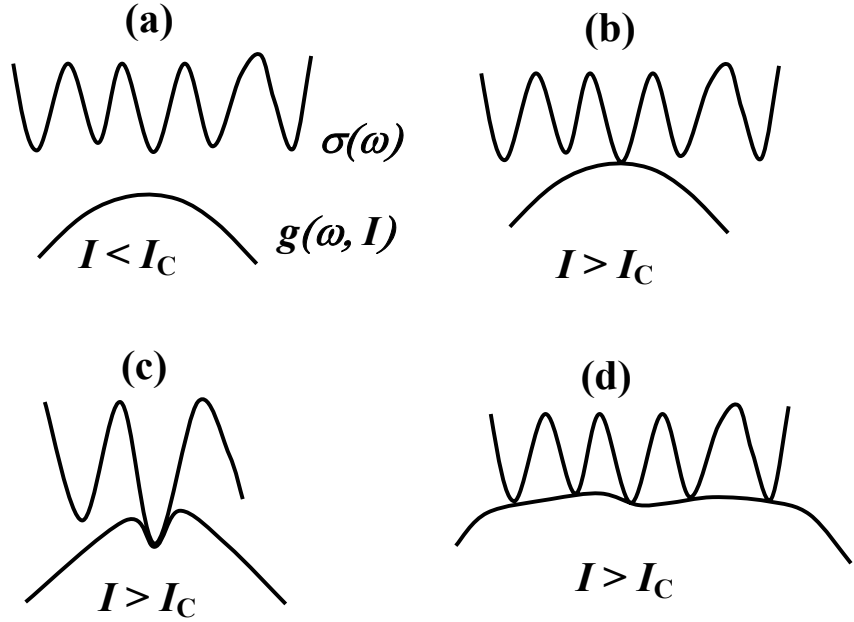

**Figure 1**. Relative configuration of the spectral curves corresponding to the material gain $g(\omega, I)$ and the loss $\sigma(\omega)$ of the electromagnetic field intensity at the cavity mirrors below (a) and above (c-d) the threshold $I_C$.

## 2. Standard model of semiconductor laser

The simplest model of the laser is a pumped two-level electronic system, immersed in an electromagnetic wave resonator. It is described (see, e.g. [**4**]) by rate equations for the electron population difference equation $n = n_2 - n_1$ and the number of photons $N$ in the resonator:

$$\frac{dn}{dt} + \frac{n(t) - n_0}{\tau} = I - \gamma n N$$
$$\frac{dN}{dt} + \sigma N = \gamma n N \qquad (1)$$

where the differential gain $\gamma$, defined by $g(\omega, I) = \gamma(\omega) n(I)$, is a coefficient independent of $n$ and $\tau$ is the characteristic time describing all non-stimulated recombination processes (in high quality material, where non-radiative

- 2 -

recombination is negligible, $\tau = \tau_{sp}$, where $\tau_{sp}$ is the characteristic time of spontaneous emission). The equilibrium population difference at $I = 0$ is denoted by $n_0$. Laser generation begins when the photon gain $\gamma n$ exceeds loss $\sigma$. In this case, the stationary solution of Eq. (1) is $\gamma n = \sigma$ and $N = (I - I_C)/\sigma$, where $I_C \equiv (\sigma/\gamma\tau) - n_0/\tau$.

In this simplest model, the $I$ dependence of gain $g(\omega, I)$ is parameterized by a single number $n$ and the rigidity illustrated in Fig. 1b arises automatically. Above the threshold, the mean-field equations (1) describe a wide range of phenomena, including relaxation of arbitrary initial state to the steady state at given $I$.

The standard STLH theory of laser linewidth is developed as follows. In the limit $N \gg 1$, the electromagnetic field $\widetilde{E}(t) = E \exp(i\omega_0 t)$ of a single resonator mode is considered classical, characterized by amplitude and phase. Here $\omega_0$ is the mode frequency, and $E$ is a complex vector that may be slowly varying in time. In the mean-field approximation, the phase $\varphi$ of the field is definite, while its amplitude is proportional to $\sqrt{N}$, i.e. $E \sim \sqrt{N} \exp(i\varphi)$. Beyond the mean-field approximation the quantities $N$, $n$, and $\varphi$ fluctuate in time due to the randomness of recombination and relaxation processes. It is these fluctuations that determine the linewidth in the conventional STLH approach. In an idealized laser, the fluctuations arise from randomness of spontaneous emission. All fluctuations of interest, including spontaneous emission, can be described classically in the sense that they are generated by δ-correlated Langevin forces (white noise). The reason for the classical description of fluctuations is that the time scale we are interested in (of order the inverse linewidth) is long compared to all kinetic relaxation times.

In the limit $N \gg 1$, where the fluctuations in the number of photons are small, $\delta N \ll N$, the fluctuations of $\varphi$ are decoupled from those of $n$ and $N$. Fluctuations $\delta N$ and $\delta n$ give rise to the intensity noise, while only fluctuations of the phase, $\delta\varphi$, contribute to the linewidth. These fluctuations correspond to a random walk of the complex variable $E$ of a constant modulus (see e.g. [**3**]). Each event of spontaneous emission adds to vector $E$ a small $\delta E \sim \sqrt{\hbar\omega_0}$. The angle between the two complex numbers $E$ and $\delta E$ is random and both the amplitude and the phase of the sum $E + \delta E$ are varying. The amplitude variation, $|E + \delta E|^2 - |E|^2$, corresponds to $\delta N$ and, according to Eq. (1), it relaxes to its steady-state value, while $\delta\varphi \approx \delta E / E \approx 1/\sqrt{N}$. The diffusion coefficient describing the angular random walk, $D_\varphi = (\delta\varphi)^2 / \tau_{sp}$, determines the laser linewidth, $\Gamma = D_\varphi$, which thus turns out to be inversely proportional to the intensity of laser emission,

$$\Gamma_{STLH} = \frac{1}{\tau_{sp} N} \ . \tag{2}$$

Thus, at large $N$, the linewidth is much smaller than any characteristic frequency of the system, such as the spectral width of the laser cavity $\sigma(\omega)$, the rate of electronic collisions $1/\tau_{ee}$ that determine the broadening of the quantum electronic levels in semiconductors, the spectral width of the gain $g(\omega)$, and the spontaneous emission[1] rate $1/\tau_{sp}$.

We would like to stress that the STLH approach essentially relies on the assumption that the mean-field equations have a singular solution with no width at all. Discussion of this assumption requires a detailed analysis of the injection-level dependence $g(\omega, I)$, which in turn requires a consideration of energy and frequency dependences of the electron and photon distributions, $n_\varepsilon$ and $N_\omega$, respectively. In Sect. 3 we discuss such a description based on Boltzmann's kinetic equation. It turns out that singular solutions are ruled out in the kinetic description which yields a finite laser linewidth already in the absence of fluctuations.

## 3. Kinetic equation

The simplest kinetic equation describing the energy distribution of electrons $n_\varepsilon$ and photons $N_\omega$ is of the form

$$\frac{dn_\varepsilon}{dt} = -\gamma_\varepsilon n_\varepsilon N_\omega + I_\varepsilon + S\{n_\varepsilon\} \tag{3a}$$

$$\frac{dN_\omega}{dt} = \gamma_\varepsilon n_\varepsilon N_\omega - \sigma_\omega N_\omega \tag{3b}$$

where the energy parameters $\varepsilon$ and $\omega$ are related by $\omega = \varepsilon(k) + E_G$, with $E_G$ being the bandgap energy and $\varepsilon(k) \equiv \varepsilon_e(k) + \varepsilon_h(k)$ the kinetic energy of carriers at a wavevector $k$ corresponding to the transition. In terms of the dimensionless $n_\varepsilon$, the total electron population difference $n$ that enters Eq. (1) can be expressed as

$$n = \int_0^\infty n_\varepsilon \nu(\varepsilon) d\varepsilon,$$

where $\nu(\varepsilon)$ is the density of electronic states. Similarly, the total injection level is $I = \int I_\varepsilon \nu(\varepsilon) d\varepsilon$, where $I_\varepsilon$ is the differential injection intensity.

---

[1] The precise meaning of the spontaneous emission rate is unclear in this model. The question is what is the spectral width for spontaneous emission events that appear in the derivation of Eq. (2)? For example, in some scenarios one may take into account only the spontaneous emission into the linewidth $\Gamma$ itself, in which case the power of the pumping intensity in Eq. (2) would be different.

The collision integral $S$ comprises contributions from electron-electron, electron-phonon interactions, and non-stimulated recombination,

$$S\{n_\varepsilon\} = S_{ee} + S_{e-ph} + S_{rec}. \qquad (4)$$

We consider the simplest situation when the electron-electron scattering rate $1/\tau_{ee}$ is fastest. This situation is also most relevant for semiconductor lasers operating at room temperature. The collision integral $S_{ee}$ is nullified by the Fermi distribution function $n_\varepsilon^F$, parameterized by an arbitrary chemical potential $\mu_{eff}$ and temperature $T_{eff}$. These parameters are determined from the conservation laws for the number of particles and energy, which can be obtained from (3) by integrating over $\varepsilon$ and $\omega$. At room temperature, the energy relaxation rate is fast and one has $T_{eff} = T$.

The distribution function $n_\varepsilon$ deviates from the Fermi shape in a narrow interval of energies of order the linewidth $\Gamma$, where $N_\omega \neq 0$ and $n_\varepsilon \equiv n_\varepsilon^F + \delta n_\varepsilon$. The typical energy exchange involved in electron-electron scattering events is of the order of $T$ and in the limit $\Gamma \ll T$ the relaxation time approximation for electron-electron scattering is exact,

$$S_{ee} = -\delta n_\varepsilon / \tau_{ee}. \qquad (5)$$

The reason for this is that $\delta n_\varepsilon$ in region $\Gamma$ is formed by incoming and outgoing fluxes from a much larger region of order $T_{eff}$ or $\mu_{eff}$ (whichever is larger). According to Eq. (3b), in a stationary state ($dN_\omega/dt = 0$) the electron distribution function is pinned in region $\Gamma$ and is independent of the injection level $I$ or its energy distribution $I_\varepsilon$:

$$n_\varepsilon = \sigma_\omega / \gamma_\varepsilon. \qquad (6)$$

On the other hand, the electronic distribution in the region outside $\Gamma$ (where $N_\omega = 0$) is not pinned because the escape rate from the outside region into the active region $\Gamma$ is finite and characterized by a time constant of order $\tau_{ee}$. The total electron concentration outside region $\Gamma$ hence grows with the injection $I$. This means that the width of $\Gamma$ itself increases with $I$.

To make this argument quantitative, we note that $\delta n_\varepsilon$ vanishes at the edges of region $\Gamma$. Depending on the shape of the function $f(\varepsilon) = \sigma_\omega / \gamma_\varepsilon$ in the right-hand side, Eq. (6) may have many solutions which correspond to the existence of multiple lasing modes in the mean-field approximation. Let us focus on the single-mode case, when $f(\varepsilon)$ has a single minimum at $\varepsilon = \varepsilon_0$ and is approximated by $f(\varepsilon) = f(\varepsilon_0) + a(\varepsilon - \varepsilon_0)^2$, where $f(\varepsilon_0) \approx 1$, see Fig. 2. The shape of $f(\varepsilon)$ can be characterized by a halfwidth, $\Delta \approx 2\sqrt{1/a}$. In the case when $\sigma_\omega$ is a sharper function than $\gamma_\omega$, the quantity $\Delta$ is the resonator linewidth. Within the interval $\Gamma$ we can write

$$\delta n_\varepsilon = \tfrac{1}{4} a \Gamma^2 - a(\varepsilon - \varepsilon_0)^2 \qquad (7)$$

where the constant is chosen so that $\delta n_\varepsilon = 0$ for $\varepsilon - \varepsilon_0 = \pm \tfrac{1}{2}\Gamma$.

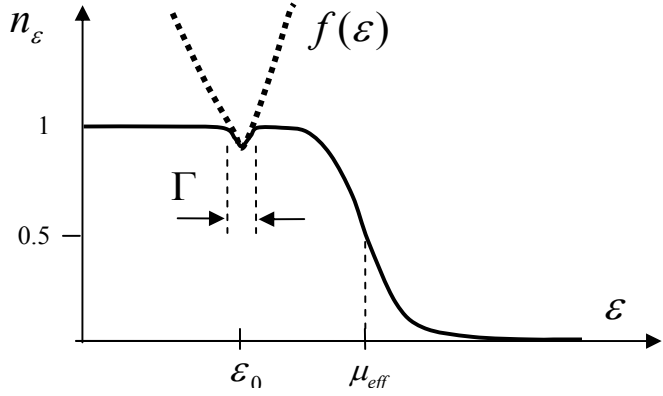

**Figure 2.** Schematic representation of the electron energy distribution $n_\varepsilon$ and the function $f(\varepsilon) \equiv \sigma_\varepsilon / \gamma_\varepsilon$. These functions coincide in region $\Gamma$.

Integrating Eq. (3a) over all energies in the stationary case ($dn_\varepsilon / dt = 0$) we find

$$I - I_C = \int_0^\infty \sigma(\omega) N(\omega)\, d\omega \qquad (8)$$

where the threshold injection $I_C$ equals

$$I_C = -\int_0^\infty S_{rec} \nu(\varepsilon)\, d\varepsilon \qquad (9)$$

(terms $S_{e-ph}$ and $S_{ee}$ drop out when integrated over all energies since they conserve the number of electrons). We note that the integrand in (8) is nonvanishing only in the small region $\Gamma$ that is much narrower than either the effective temperature $T_{\text{eff}}$ or the Fermi level $\mu_{\text{eff}}$. Therefore, if we integrate Eq.(3a) over $\Gamma$, we find

$$I - I_C = -\int_\Gamma \frac{\delta n_\varepsilon}{\tau_{ee}} \nu(\varepsilon) d\varepsilon . \qquad (10)$$

Substituting Eq. (7) into Eq. (10) we obtain an estimate of the laser linewidth:

$$\Gamma^3 = \frac{6}{a\nu(\varepsilon_0)} (I - I_C) \tau_{ee}$$

or, equivalently,

$$\Gamma = \Delta \left( \frac{3}{2\nu(\varepsilon_0)\Delta} \right)^{1/3} \left[ (I - I_C) \tau_{ee} \right]^{1/3} . \qquad (11)$$

We see that the linewidth in the mean-field approximation increases with pumping. This is in drastic contradiction with the conventional STLH result (2) that predicts a linewidth decreasing with $I$.

The fundamental reason for this discrepancy is the assumption by STLH of a singular, $\delta(\omega - \omega_0)$ like, frequency dependence of the field $E(\omega)$ in the absence of fluctuations. In contrast, the solutions of kinetic equations are smooth functions of $\varepsilon$ and $\omega$ and do not exhibit any singularity. Consequently, an account of fluctuations would make only a small correction to our result.

**4. Discussion**

It should be cautioned that validity of kinetic equations (3) requires that the uncertainty in electronic energies due to collisions be smaller than the interval of electronic energies that we are interested in ($1/\tau_{ee} \ll \Gamma$). According to Eq. (11), this condition is satisfied at sufficiently high injection intensities. However, semiconductor lasers at room temperature are typically in the opposite regime $1/\tau_{ee} \gg \Gamma$. In this regime we are concerned with the details of the electron distribution function resolved on a much finer scale than that on which the single electronic states themselves are well defined. We are not aware of any example in kinetic theory where a quantitative description of such a situation has been developed. Its qualitative physical aspects can to some extent be captured in a model that relaxes the strict energy conservation in single-electron transitions,

$$\delta(\varepsilon - \omega) \Rightarrow \Theta(\varepsilon - \omega) \sim \frac{\tau_{ee}}{\tau_{ee}^2 (\varepsilon - \omega)^2 + 1} \qquad (12)$$

Although this model will give a somewhat different expression for the linewidth compared to Eq. (11), it is clear that $\Gamma$ will remain an increasing function of $I$.

Available experiments lend conclusive support neither to our result nor STLH. At low intensities above threshold one observes a decreasing linewidth but at higher intensities the linewidth often saturates and then re-broadens, so that $\Gamma(I)$ exhibits a minimum (see, e.g., Fig. 6.15 in [**5**], Fig. 9.11 in [**6**], or the more recent data [**7**]). One of the possible scenarios that would reconcile the two pictures is that at low injection, the mean-field linewidth given by the kinetic equation approach happens to be much smaller than the STLH linewidth given by (2), i.e. $\Gamma(I) < \Gamma_{STLH}(I)$ at least near threshold. In this case, the initial decrease of the linewidth with $I$ could be attributed to a STLH-like mechanism, whereas for larger $I$ the increasing mean-field linewidth takes over and one has re-broadening.

In the opposite limit, which we find more realistic, there is no range for STLH to hold and we would have to conclude that the decreasing linewidth lacks theoretical explanation. Development of a satisfactory linewidth theory would then require inclusion of additional phenomena that go beyond the kinetic description. We would like to mention here two such phenomena:

(***a***) If the spectral width of the laser oscillations is narrower than the energy width of single electron states, which is of order $1/\tau_{ee} \gg \Gamma$, then the energy conservation low should only be satisfied with the precision of $1/\tau_{ee}$. In this case, in Eq.(3a) the term $nN$, which is responsible for the electron-hole recombination rate, should be replaced by a term proportional to

$$n(t)|E(t)|^2 . \qquad (13)$$

When the electric field is monochromatic, these two expressions are identical, and we come back to Eq.(3a). However, for a finite frequency range, Eq. (13) leads to beatings between different frequency components of the field. With the electron concentration $n(t)$ exhibiting beatings in time the problem becomes non-stationary and highly nonlinear. This applies both to the case of frequency-distributed field within a single mode and to the multimode case. In the latter case a related problem arises: the dependence of the number of lasing modes on the pumping intensity. This dependence is often nonmonotonic, increasing at small $I$ and decreasing at large $I$. In a broad sense, it could be interpreted as a narrowing of the total spectral width of laser oscillations. An attempt has been made [**8**] to explain this phenomenon by a mode competition generated by the term in Eq. (13). As far as we know, however, this problem remains unsolved.

(***b***) Different harmonics of lasing radiation have different spatial dependencies. The electron recombination rate depends on the local amplitude of the electromagnetic field. Thus, different harmonics of the electromagnetic field can compete via the spatial dependence of the electron concentration $n(r)$ due to the spatial hole burning. As far as we know, the significance of this effect for the laser line width has not been elucidated.

## 5. Conclusion

We see that the standard theory of laser linewidth is unsatisfactory. The theory attributes the spectral width of laser oscillation to fluctuations brought about by random spontaneous emission events and is essentially based on the assumption that in the absence of fluctuation laser radiation is monochromatic. We have shown that this assumption is inconsistent and that already in the mean field model the laser oscillations have a finite spectral linewidth that furthermore increases with pumping.

Our consideration was restricted to semiconductor lasers but our conclusion is likely to be more general, applicable to other lasers as well, such as solid-state lasers and gas lasers. The question of why the laser linewidth can be much narrower than either the gain spectrum or the resonator linewidth is begging a theoretical explanation.

**Acknowledgement**

We are grateful to R. F. Kazarinov for useful discussions.

**References**

1. A. L. Schawlow and C. H. Townes, "Infrared and optical masers", *Phys. Rev.* **112**, pp. 1940-1949 (1958).

2. M. Lax, "Classical noise v. noise in self-sustained oscillators", *Phys. Rev.* **160**, pp. 290-307 (1967); R. D. Hempstead and M. Lax, "Classical noise VI. Noise in self-sustained oscillators near threshold", *Phys. Rev.* **161**, pp. 350-366 (1967).

3. C. H. Henry, "Theory of linewidth of semiconductor lasers," *IEEE J. Quantum Electronics* **QE-18**, pp. 259-264 (1982); "Theory of the phase noise and power spectrum of a single-mode injection laser," *ibid.* **QE-19**, pp. 1391-1397 (1983).

4. A. E. Siegman, *Lasers*, University Science Books, Sausalito, CA (1986).

5. G. P. Agrawal and N. K. Dutta, "*Semiconductor Lasers*", 2nd edition, Van Nostrand, New York (1993).

6. G. Morthier, P. Vankwikelberge, *Handbook of Distributed Feedback Laser Diodes*, Artech House, Inc., Boston (1997).

7. H. Su, L. Zhang, R. Wang, T. C. Newell, A. L. Gray, and L. F. Lester, "Linewidth study of InAs-InGaAs quantum dot distributed feedback lasers", *IEEE Phot. Technol. Lett.* **16**, pp. 2206-2208 (2004).

8. R. F. Kazarinov, C. H. Henry, and R. A. Logan, "Longitudinal mode self-stabilization in semiconductor lasers," *J. Appl. Phys.* **53**, pp. 4631-4644 (1982).